\documentclass{article}
\usepackage{smc}
\usepackage{times}
\usepackage{ifpdf}
\usepackage[english]{babel}
\usepackage{cite}

\def\papertitle{A Perceptual Study of Sound Ecology in Peripheral Sonification}
\def\firstauthor{Maxime Poret}
\def\secondauthor{Catherine Semal}
\def\thirdauthor{Myriam Desainte-Catherine}

\newif\ifpdf
\ifx\pdfoutput\relax
\else
   \ifcase\pdfoutput
      \pdffalse
   \else
      \pdftrue
\fi

\ifpdf 
  \usepackage[pdftex,
    pdftitle={\papertitle},
    pdfauthor={\firstauthor, \secondauthor, \thirdauthor},
    bookmarksnumbered, 
    pdfstartview=XYZ 
   ]{hyperref}

  \usepackage[pdftex]{graphicx}
  \graphicspath{{./figures/}}
  \DeclareGraphicsExtensions{.pdf,.jpeg,.png}

  \usepackage[figure,table]{hypcap}
  \usepackage{multirow}
  \usepackage{gensymb}

\else 
  \usepackage[dvips,
    bookmarksnumbered, 
    pdfstartview=XYZ 
  ]{hyperref}  

  \usepackage[dvips]{epsfig,graphicx}
  \graphicspath{{./figures/}}
  \DeclareGraphicsExtensions{.eps}

  \usepackage[figure,table]{hypcap}
\fi

\hypersetup{
    colorlinks,%
    citecolor=black,%
    filecolor=black,%
    linkcolor=black,%
    urlcolor=black
}

\title{\papertitle}

 \threeauthors
   {\firstauthor} {Univ. Bordeaux,\\ Bordeaux INP, CNRS, LaBRI,\\ UMR5800, F-33400 Talence, France \\ %
     {\tt \href{mailto:maxime.poret@labri.fr}{maxime.poret@labri.fr}}}
   {\secondauthor} {Univ. Bordeaux,\\ Bordeaux INP, CNRS, INCIA,\\ UMR5287, F-33076 Bordeaux, France \\ %
     {\tt \href{catherine.semal@ensc.fr}{catherine.semal@ensc.fr}}}
   {\thirdauthor} {Univ. Bordeaux,\\ Bordeaux INP, CNRS, LaBRI,\\ UMR5800, F-33400 Talence, France \\ %
     {\tt \href{myriam@labri.fr}{myriam@labri.fr}}}

\begin{document}
\capstartfalse
\maketitle
\capstarttrue
\begin{abstract}
Based on a case study on 3D printing, we have been experimenting on the sonification of multidimensional data for peripheral process monitoring. In a previous paper, we tested the effectiveness of a soundscape which combined intentionally incongruous natural and musical sounds. This was based on the hypothesis that auditory stimuli could better stand out from one another if they were less ecologically coherent, thus allowing for better reaction rates to various notifications. In this paper, we follow up on that hypothesis by testing two new acoustic ecologies, each exclusively consisting of either musical or natural sounds. We then run those ecologies through the same dual-task evaluation process as the previous one in order to compare them. The results seem to favor our hypothesis, as the new ecologies were not detected as accurately as the original. Though, the set of natural sounds seemed to be considered less intrusive by testers, and to allow for a better performance at an external primary task. We hope to see this work become part of a much larger corpus of studies, which may eventually provide a more definite answer on the effect of ecological coherence in peripheral soundscape design.
\end{abstract}

\section{Introduction}\label{sec:introduction}

The display of data through non-verbal sounds, or sonification~\cite{kramer1999}, has been shown to adequately provide various types of notifications for process monitoring tasks by Gaver et al.~\cite{gaver1991}, then Rauterberg et al.~\cite{rauterberg1994}, even for large numbers of monitoring criteria. Although historically it has been used for decades to convey alerts in industrial processes (often in the overly-stressful form of bells, whistles, honks and beeps), the advent of more advanced sound and data processing technologies in recent years has brought on an effort to move on from the ``better safe than sorry'' paradigm~\cite{patterson1990} towards calmer, less invasive soundscapes~\cite{weiser1996}. In this paper, we focus on the ``peripheral'' kind of sonification, which should give listeners access to various auditory notifications on data states while also allowing them to take care of more attention-demanding primary tasks~\cite{vickers2011}.

Traditionally, the design guidelines for such sonifications recommend using a small collection of concurrent, continuously playing sound streams~\cite{hildebrandt2016}, which should be selected with ecological coherence in mind. This means both making sure that the streams are not masked by sounds otherwise present in the use context, and ensuring that the spectral bandwidths of the sounds are allocated in a way that prevents them from masking one another~\cite{vickers2011, rauterberg1994, mynatt1998}. Our goal here is not to question those precautions against masking, as they seem quite self-evident. However, sonification designers often appear to follow an extra rule that this coherence should also involve a particular attention to logical likelihood in the sound choices, so that soundscapes usually revolve around themes (or ``ecologies'') such as ``forest'', ``beach'', or various forms of exclusively musical displays.

Recent research into auditory processing suggests that the discrimination of sounds happens on a neurophysiological level through a hierarchy of parallel classifications~\cite{murray2006, lomber2008, leaver2010}, and that natural sounds, even highly shortened and simplified, can still be recognized amidst incongruous sets of musical sounds~\cite{isnard2016}. We wonder, then, if it might be advantageous to select sounds belonging to very different cognitive categories (e.g. musical sounds vs. animal cries) so that cortex-level classification processes may better separate them.

In a previous work on process monitoring sonification~\cite{poret2021}, we formulated the hypothesis that an incongruous set of stimuli might allow for a faster and more accurate discrimination of notifications than a logically coherent one. Our goal with this work is to conduct an experiment testing this hypothesis.

It is quite common for researchers studying process monitoring sonification to propose several sets of stimuli based on different design strategies (often a dichotomy of auditory icons versus musical mappings)~\cite{cohen1993, gilfix2000, hermann2015, lenzi2019, aldana2020}. However, works including a ``mixed'' ecology of both musical and natural sounds seem to be more difficult to come by \cite{mynatt1998, matinfar2019} and none, to our knowledge, has produced a full comparative evaluation of those design choices beyond qualitative user surveys.

In this work, we designed and evaluated two ecologically coherent sets of stimuli in order to compare them to an incoherent one we previously tested. To this end, we reused some of the existing sound streams and completed them with new ones, which were selected according to the respective ecological guidelines of the new sets (Section~\ref{sec:soundmaps}). We then evaluated them using the same dual-task experiment as before (Section~\ref{sec:expe}) so that we could compare, for each ecology, the quality of anomaly annotation, the effects on primary task performance, and the listeners' aesthetic appreciations (Section~\ref{sec:results}).

\section{Context}
\subsection{3D Printing}
This project stems from a case study on the use of sonification to help monitor the wire arc additive manufacturing process\cite{williams2016}. During this process, metallic parts are built up layer by layer by the addition of material which gets heated up, melted and deposited by bolts of high-voltage current running through a wire. As such a process tends to be noisy, flashy, and overall unpleasant to be around (let alone monitor directly), there has been an effort in the past few years to augment its observation through the use of alternative modalities, such as haptic vibrations~\cite{ibarboure2021}, visual augmented reality~\cite{Ceruti2017} or, in our case, a peripheral auditory display~\cite{poret2020,poret2021}.

The peripheral aspect of this display is expected to provide operators with a more remote form of monitoring, which would allow them to move away from the printer and take care of other tasks such as checking their e-mail, managing their schedules, handling other machines, etc., while still being able to come back to the process in reaction to notifications.

The manufacturing criteria that our display should help monitor are described below. See also the first column of Table~\ref{tab:maprecap} for the full detail of thresholds and tolerances.
\begin{itemize}
\item The Weld Pool Dimensions (WPD) are the width and height of the material deposit right below the printing head.
\item The Part Height (PH) is the cumulative height of all the constructed layers below the printing head.
\item The Weld Pool Temperature (WPT) is the local temperature of the material deposit.
\item The Part Temperature (PT) is the average temperature of the part being constructed.
\end{itemize}

\subsection{Simulated Process}
As of writing this article, the sensors which would be able to provide those monitoring data are still being researched\cite{xia2020}. Instead, our sonification is produced by running SuperCollider\footnote{https://supercollider.github.io/} scripts on synthetic log files which emulate the way criteria should behave during the printing process. 

In order to properly evaluate the sounds-to-criteria connections, one would need to account for users' learning curve over a proper training phase, so the experiment would need to take place on a relatively long term (several trials a week over a month or two). This would also ideally require a population of printer operators, who would already be well-acquainted with the printing process and familiar with its conditions and criteria.

In the current state of this project, such in-situ experimentation is not yet feasible. So, we decided not to concern ourselves with studying the mental connections between sound metaphors and manufacturing criteria yet, and to instead focus our experimental process on the detection of the sounds themselves, as well as their usability in a (simulated) peripheral work context. Despite this limitation, we hope to lay the groundwork for future larger-scale experiments studying the use of similar displays in real monitoring situations.

\section{Sound mappings}\label{sec:soundmaps}

\begin{table*}[t]
\centering
\def\arraystretch{1.2}
\begin{tabular}{|c|c|c|c|}
\hline
    \multirow{6}{*}{\shortstack{WPD\\ Width: $4mm \pm 10\%$\\ Height: $3mm \pm 10\%$}} & Height difference & \shortstack{Tonic pitch: C5 (normal) to F6 (worst)} & \multirow{3}{*}{\shortstack{Arpeggio (Mixed \& Synth)}} \\\cline{2-3}
     & Both differences & \shortstack{Loudness: 0.02x to 0.2x} &  \\\cline{2-3}
     & Width difference & \shortstack{Inter-offset interval: 0.5 to 1.5 s} &  \\\cline{2-4}
     & Height difference & \shortstack{Playback speed: 0.5x to 2.0x} & \multirow{3}{*}{\shortstack{Droplets (Nature)}} \\\cline{2-3}
     & Both differences & \shortstack{Loudness: 0.1x to 0.8x} &  \\\cline{2-3}
     & Width difference & \shortstack{Intervals: 0.5 to 1.5 s} &  \\\hline
    \multirow{4}{*}{\shortstack{PH\\Current layer $\pm 1.5mm$}} & Relative difference & Pitch: A3 $\pm 3$ tones & \multirow{2}{*}{Drone (Mixed \& Synth)} \\\cline{2-3}
     & Absolute difference & Loudness: 0.1x to 0.4x & \\\cline{2-4}
     & Relative difference & \shortstack{Mixing: Ducks, Misc. or Crows} & \multirow{2}{*}{Birds (Nature)} \\\cline{2-3}
     & Absolute difference & Loudness: 0.0x to 0.3x &  \\\hline
    \multirow{4}{*}{\shortstack{WPT\\$2000 \degree C \pm 10\%$}} & Difference polarity & Pitch: 220 Hz or 880 Hz & \multirow{2}{*}{Jingle (Synth)} \\\cline{2-3}
     & Absolute difference & Loudness: 0.0x, 0.1x or 0.2x &  \\\cline{2-4}
     & Difference polarity & Selection: Crackling or Boiling & \multirow{2}{*}{Water (Mixed \& Nature)} \\\cline{2-3}
     & Absolute difference & Loudness: 0.0x, 0.2x or 0.5x &  \\\hline
    \multirow{2}{*}{\shortstack{PT\\$600\degree C$}} & \multirow{2}{*}{Threshold} & Trigger: Random components & Bell (Synth) \\\cline{3-4}
     &  & Trigger: Sound sample & Sizzle (Mixed \& Nature) \\
     \hline
\end{tabular}
\caption{Overview of the mappings from each monitoring criterion (left) to each stimulus (right). In parentheses after each stimulus name: the names of the ecologies in which it is involved.}
\label{tab:maprecap}
\end{table*}

\subsection{General Principles}

Even though the association of sounds to criteria is not studied here, we want to be prepared for that next step by carefully selecting metaphors that relate to the different types of physical data.

For our natural stimuli, we avoided selecting concurrent sounds that would be too difficult to differentiate without any visual context. Indeed, for example, the sounds of heavy rain, boiling water, or a flowing river, may easily be confused for one another, or mask each other when occurring at the same time. Likewise, when selecting our musical stimuli, we made sure that the streams had non-overlapping pitch ranges, as well as different timbres and granularities.

The choices of timbre, instrument, or sound source vary from ecology to ecology, but we aim to maintain as many psychoacoustic concerns as possible across all ecologies. The thematic dimension is the main variable here. As such, the soundscape always has an ``idle'' state made up of two continuous stimuli indicating normal WPD and PH values. Both dimensions for WPD are mapped to three parameters (pitch for height, rhythm for width, and loudness for both), defining a slowly-repeating burst of sound, which metaphorically represents the progressive accumulation of material. PH is conveyed by a continuous sound, which can evolve into three states (normal, too low, too high) and represents the base upon which the material is being deposited. WPT is conveyed by silence in normal situations, but can emerge in one of two forms depending on the direction of the anomaly. Indeed, there is no need to constantly remind listeners that the temperature is right, especially if we want to avoid overloading the display in the absence of anomalies. PT is conveyed by a single sound event denoting the passing of the threshold. It is a last-resort alarm which we choose to make conceptually related to the sound for WPT, as the global temperature alarm is expected to be brought on by a prolonged local temperature anomaly.

The following subsections give a description of the mapping choices for each ecology. See Table~\ref{tab:maprecap} for an overview of those mappings for each stimulus and how they combine to form each of the three ecologies.

\subsection{Mixed Ecology}\label{subsec:Mixed}
As previously described in~\cite{poret2021}, the Mixed ecology was designed so that geometrical (WPD, PH) and thermal (WPT, PT) criteria would be conveyed by two distinct categories of sounds, respectively based on auditory icons~\cite{gaver1993} and abstract ear\-cons~\cite{blattner1989} guidelines.

The thermal criteria were relatively easy to convey through metaphorical auditory icons, as there are many real-world phenomena whose sounds directly relate to the idea of temperature (boiling water, crackling ice, burning fire, sizzling water). The geometrical criteria, however, required further levels of abstraction in order to be conveyed by sounds. Indeed, it is rather difficult to directly correlate natural phenomena to the concepts of ``width'' or ``height''. For this reason, we instead chose to map those dimensions to perceptual parameters in musical streams.

The Mixed ecology thus consists of the 4 stimuli Arpeggio (WPD), Drone (PH), Water (WPT) and Sizzle (PT), as detailed below.

Arpeggio is a looping motive of 3 pitches (tonic, third and fifth) in the major scale, played by the default SuperCollider synth, a simplistic piano-like sound. In idle state, this sound stream consists in a repetition of a low-volume C5 Major arpeggio, with each note separated by 1.5 seconds. When the tonic pitch, duration, or loudness is changed by a fluctuation in WPD (see mappings in Table~\ref{tab:maprecap}), the newly-defined motive interrupts the one previously started.

Drone is a continuous bandpass-filtered sawtooth wave playing a fluctuating pitch. In idle state, it plays a constant low-volume A3 note. From there, its pitch can get lower or higher following the direction of the PH discrepancy.

Water can be seen as a small body of water which is idle and silent most of the time, but can start boiling or freezing as a result of unnaturally quick changes in temperature.

As a continuation of this Water metaphor, the Sizzle alarm consists in a more violent evaporation sound, reminiscent of water being poured onto a red-hot surface.

\subsection{Synth Ecology}
For the Synth ecology, we only chose musical sounds, using a variety of synthetic timbres with abstract mappings to perceptual parameters. We reused Arpeggio and Drone as defined for Mixed and designed two new stimuli for thermal dimensions WPT and PT.

For WPT, the natural metaphor selection was replaced by a binary mapping to the pitch of a rapid percussive sound grain (a repeating 0.06s sine wave burst with quick attack and decay), which we call ``Jingle''. This stimulus is silent in idle state. It emerges in loudness as an anomaly arises, and its pitch can take one of two values (220 Hz for under-heating and 880 Hz for overheating).

The PT threshold alarm was replaced by ``Bell'', an inharmonic sound with quick attack and slow decay reminiscent of a large bell being struck. Like previously with Sizzle and Water, this metaphor was chosen as a continuation of the Jingle timbre. We boosted this alarm's novelty aspect for listeners by making its timbre partially randomly-generated: around a constant 440 Hz main frequency, 3 inharmonic components are randomly selected between 220 and 880 Hz.

\subsection{Nature Ecology}
The Nature ecology is based on the principles of auditory icons design. We reused the Water and Sizzle auditory icons previously defined for Mixed and added two new stimuli for geometrical dimensions WPD and PH, using metaphors akin to a ``pond in the forest'' ecological theme.

The Nature ecology thus consists of the stimuli Droplets (WPD), Birds (PH), Water (WPT) and Sizzle (PT).

As noted in the description for the Mixed ecology, it is difficult to select natural sound metaphors which can be related to the geometrical dimensions of an item. But we can benefit from a property of natural sound processing, which Gaver calls ``everyday listening''~\cite{gaver1989}, by which listeners detect, from the perceptual properties of a sound, the characteristics of the object or action producing it. For instance, a larger object falling into a pond produces a lower-pitched and longer ``splash'' sound than a smaller one. It is by a similar logic that we defined the Droplet stimulus, which conveys WPD using a short sound sample of a drop of water falling onto a surface. The height was mapped to its playback rate so that it would sound higher-pitched if the weld pool became higher (as if falling harder), and the width was mapped to its occurrence rate so that it would be heard more often as the weld pool became wider (as if overflowing).

The Birds stimulus conveying PH is based on the idea of constructing a continuous layer of miscellaneous birdsong, such that in idle state no information stands out from an overall pleasant and familiar noise. In case of an anomaly, a single species sound emerges, thus breaking the noise and calling for the listener's attention. That species is either a duck if PH is too high or a crow if it is too low.

This abstract metaphorical choice of ducks or crows to translate height is an attempt to map that dimension to the respective affective connotations of those birds in Western folklore: the duck is a familiar domesticated bird (positive opinion, so a ``high'' polarity), while the crow is a scavenger and a bad omen (negative opinion, ``low'' polarity).

\section{Experiment}\label{sec:expe}

\begin{figure}[h]
	\centering
    \includegraphics[width=0.476\textwidth]{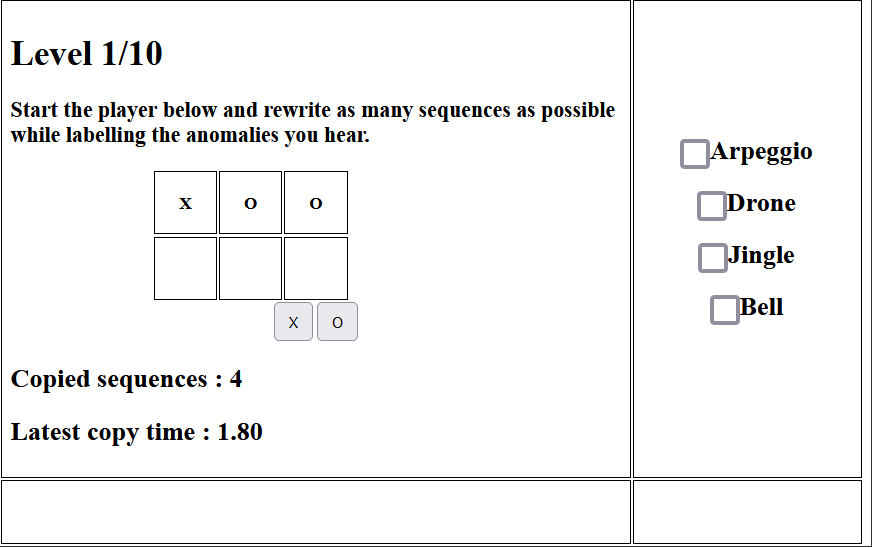}
    \caption{A screen capture of the experiment interface during a level, here for the Synth ecology. In the middle, the player rewrites the sequence displayed on the upper row by clicking the 'X' and 'O' buttons in the same order. Upon each sequence completion, a new one appears. Boxes on the right allow the player to label anomalies as they occur. The labels on these boxes depend on the names of the stimuli in use for the ecology being tested.}
    \label{fig:evalinterf}
    \end{figure}

In order to preserve the experimental conditions for comparability sake, and since it provides a basic simulation of a peripheral monitoring work context, we reused the same dual-task game that was originally built to evaluate the Mixed ecology. See Figure~\ref{fig:evalinterf} for a screenshot of the experimental interface. The primary task consists in a very simple sequence-copying task using the mouse, while the secondary task consists in listening for notifications in the soundscape and annotating them as they occur by checking boxes on the side.

Due to sanitary constraints at the time of the experiment, the evaluation game had to be sent out in the form of an interactive web page using JavaScript and PHP elements to log user actions during testing. Each participant took the test in their own environments and on their own setups, and as we acknowledge this potential lack of uniformity, we hope it can be compensated for by a large enough population sample of testers. Both sides of the experiment, featuring either one of the new ecologies, are implemented on the same page such that, upon user connection, one or the other set of stimuli is selected depending on which one has been the most represented in completed tests so far. If both ecologies have been tested the same number of times, one is picked randomly.

Participants start the experiment by undergoing a training phase during which they are introduced to the primary task, then each sound stimulus in isolation, then a few examples of the full soundscape in idle and anomalous states. Finally they have to successfully complete two levels ensuring they understood the process before moving on to the actual experiment, which consists of ten 30-seconds levels featuring various anomaly combinations selected in a random order.

The application logs the times at which anomaly boxes are checked over the course of a level, as well as the size and time of completion of each sequence copied.

Upon completing the experiment, participants are redirected towards a brief survey, first asking them for the usual demographic information (age and gender) before presenting them with a series of statements regarding the exercise (e.g. ``the task was hard'', or ``the sound was stressful'', see Table~\ref{tab:surveyAnswers}) to which they can reply with ``Disagree'', ``Somewhat agree'', and ``Agree''. Finally, a free-form comment box allows them to further develop their feedback.

\section{Results}\label{sec:results}
Synth was tested by 15 participants. Four of them had taken part in an earlier iteration of the experiment. 8M, 5F, 2 not specified. Ages 20 to 69 ($\mu = 35.9, \sigma = 15.9$).

Nature was tested by 16 participants. Two of them had taken part in a previous iteration of the experiment. 7M, 9F. Ages 19 to 62, ($\mu = 24.75, \sigma = 10.5$).

In the previous experiment, Mixed had been tested by 43 participants. 20M, 23F. Ages 18 to 67 ($\mu = 32.3, \sigma = 12.4$).

We used the data collected during the experiment to compute an array of quality metrics for the displays, detailed in the following subsections: the annotation accuracy for each stimulus, the reaction time needed to label anomalies, the distribution of primary task performances, and the qualitative user feedback.

\subsection{Annotation Accuracy}

\begin{figure}[h]
\centering
\includegraphics[width=0.47\textwidth]{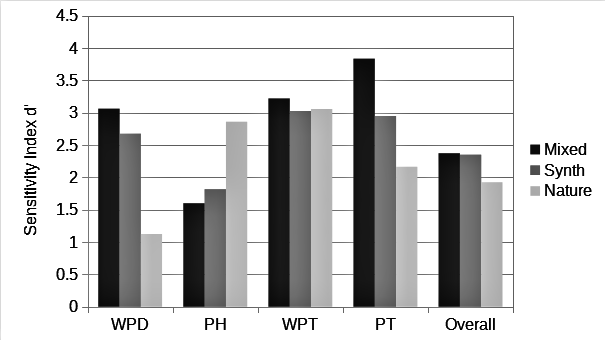}
\caption{Mean sensitivity index for each stimulus type (horizontal sections) and for each ecology (colors).}
\label{fig:sensitivityHistogram}
\end{figure}

Signal Detection Theory~\cite{macmillan2004} allows us to compute a metric called sensitivity index, or $d'$, using the hit ($H$) and false alarm ($FA$) rates of each participant in the annotation task, such that $d' = Z(H) - Z(FA)$, with $Z$ the inverse function of the cumulative normal law distribution. We applied the ``mean sensitivity'' strategy by averaging all the participants' $d'$ indices for each stimulus type. Since some users' rates were too ``exact'' to allow for $Z$ computation (total rates of 0 or 1), we approximated them to $10^{-2}$ of their actual values (respectively resulting in 0.01 and 0.99). This is equivalent to considering that, if given a hundred trials for each stimulus, the most efficient testers may still have a 1\% chance of making a mistake.

This sensitivity index is such that a higher value denotes a stronger intelligibility for the corresponding stimulus. Since it has no dimension or scale, it cannot give an absolute indication of the quality of a signal (beyond the fact that a signal with $d'>1$ has above-random chances of being correctly identified), but it can be used to compare different signals evaluated in the same experiment. 

We only considered an annotation to be a hit if its checkbox was clicked after the onset of the anomaly and remained checked until the end of the level. Each ``prediction'' of an anomaly is considered a false positive and each ``change of mind'' is counted as no annotation.

See Figure \ref{fig:sensitivityHistogram} for a histogram of the mean sensitivity index for each stimulus, as well as overall, in each ecology.
\begin{itemize}
\item WPD: Droplets is clearly less efficient than Arpeggio. Arpeggio is slightly better in Mixed than Synth.
\item PH: Birds is clearly more efficient than Drone.
\item WPT: all three ecologies do not present a clear difference, though Water appears to work slightly better in Mixed than in Nature.
\item PT: Sizzle is detected much more efficiently in Mixed than in Nature.
\item Overall: Mixed is slightly better than Synth, and Nature is visibly less efficient than either.
\end{itemize}
 
\subsection{Annotation Times}
 
\begin{figure}[h]
\centering
\includegraphics[width=0.47\textwidth]{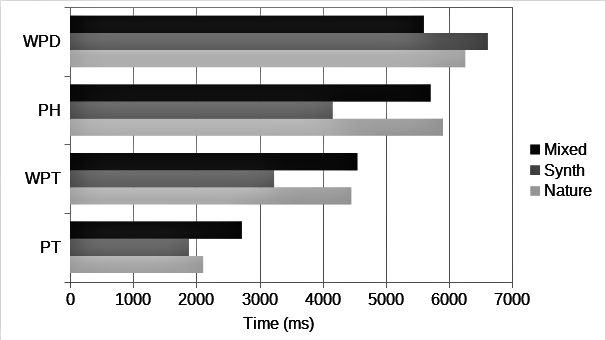}
\caption{Average annotation times in milliseconds for each stimulus type (vertical sections) and for each ecology (colors).}
\label{fig:reacTimes}
\end{figure}

In Figure~\ref{fig:reacTimes}, we plotted the average annotation times for each stimulus type and for each ecology. Once again, these are computed only from cases when participants correctly identified an anomaly after its onset. Comparing those values, we notice that Synth seems to have allowed for better reaction times overall, except for WPD, where the Arpeggio stimulus was annotated faster as part of Mixed than Synth.
 
\subsection{Primary Task Performance}
\begin{figure}[h]
\includegraphics[width=0.46\textwidth]{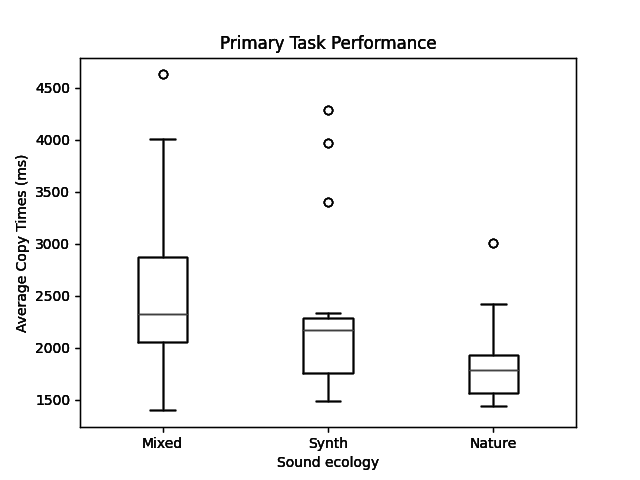}
\caption{Box-and-whiskers plots of the testers' average sequence-copying time (in ms) for each ecology.}
\label{fig:seqTimes}
\end{figure}

We computed the average time required by each participant to copy a sequence as a metric for their performance at the primary task. By plotting the distribution of those scores as box plots for each ecology (see Figure \ref{fig:seqTimes}), we get a representation of the overall performance of testers in each condition. In those diagrams, a lower value indicates a better score, since the copying time needs to be as low as possible for an optimal performance.

Nature seems to have the overall best set of participant performances, followed by Synth. The scores are more widely distributed for Mixed, since that experiment got a much larger number of participants. It seems to indicate slightly worse primary task performance than the other two.

 \subsection{Qualitative Survey}
 
 \begin{table}[h]
     \centering
     \begin{tabular}{|c|c|c|c|}
        \hline
          & Mixed & Synth & Nature \\
          \hline
         Easy to distinguish sounds & 71.43 & 82.14 & \textbf{83.33} \\
         \hline
         Sounds distract from task & 7.14 & \textbf{25.00} & 14.29  \\
         \hline
         Sounds are stressful & 19.05 & \textbf{42.86} & 30.00 \\
          \hline
         Task is stressful & 16.67 & \textbf{25.00} & 16.67 \\
          \hline
         Task is fun & 66.67 & 46.43 & \textbf{73.33} \\
          \hline
         Task is difficult & 14.29 & \textbf{21.43} & 13.33 \\
          \hline
         Task distracts from sounds & 11.90 & \textbf{21.43} & 3.33 \\
          \hline
     \end{tabular}
     \caption{Overview of the qualitative survey answers. For each statement, the strongest agreement rates are highlighted in bold font.}
     \label{tab:surveyAnswers}
 \end{table}

The end survey for the evaluation consisted in a series of seven statements (3 regarding the sound design and 4 regarding the primary task in relation to the sound), to which participants could respond on a 3-step scale from ``Disagree'' to ``Agree''. In order to analyse the answers, we attributed values to each step of that agreement scale, such that ``Disagree'' was worth 0, ``Somewhat Agree'' was worth 50, and ``Agree'' was worth 100. We then averaged those values for each statement, which gave us a collection of percentages indicating the overall agreement rates. See Table~\ref{tab:surveyAnswers} for an overview of those answers.

Overall, Synth appears to be considered more stressful than the other two ecologies (rows 2-7). This even includes testers seeing the primary task as more stressful when presented with this set of sounds (row 4).

In a free-form comment for the Synth ecology, one tester reported finding the Drone stimulus too sensitive. That issue had also been reported in the original Mixed evaluation. It seems to indicate that most of the mistakes for Drone may be caused by false alarms rather than missed hits.

The participants who tested Nature seemed to perceive its set of stimuli as easier to distinguish (row 1), despite a worse performance at the annotation task. This may be due to the fact that, although the types of sounds themselves were easy to differentiate, their modulations in case of anomalies may have been more difficult to detect. Two participants reported that they found normal and abnormal situations to be difficult to differentiate for Droplets, which is clearly reflected in its performance and was most likely encountered by more users than just the two who reported the issue.

For Nature, four testers reported having trouble differentiating Sizzle and Water, another issue which had previously been reported for Mixed. This confusion is surprising to an extent as Sizzle was selected to occupy a higher frequency range than Water, and to have a very characteristic transient envelope instead of being a continuous bubbling sound. Although those sounds were supposed to be quite different on a purely perceptual level, it seems that the ``everyday listening'' may have taken over in peripheral monitoring, and both were simply heard as ``the sound of boiling water''. Also, since it was a last-resort alarm, Sizzle was more rarely represented than other stimuli in the experiment, which may have led to testers forgetting it after the original exposure during training. We suspect that a proper long-term training may greatly reduce the problems with discriminating Water and Sizzle. Also, we note that these two stimuli still yielded better sensitivity indices in Mixed than any of the other stimuli.

The sounds used in Mixed were rated relatively more often as difficult to differentiate (row 1). This may be due to the combined effects of the aforementioned Sizzle/Water confusion and Drone over-sensitivity.

We notice that the ecologies involving natural sounds tended to be rated more positively:\begin{itemize}
    \item Nature stimuli are seen as easier to distinguish (row 1).
    \item Mixed is the least distracting and the least stressful (rows 2 and 3).
    \item Mixed and Nature are ex-aequo for least stressful influence on primary task (row 4).
    \item The primary task is seen as more fun and less difficult with Nature playing (rows 5 and 6).
    \item Nature is easier to follow while taking care of the primary task (row 7).
\end{itemize}

These last few points appears to confirm the usual arguments for using auditory icons in monitoring displays.

Finally, three participants for Nature reported that they found the sound of crackling ice used for Water to be either distracting, distressing, or stressful. This issue had not previously been reported for Mixed, and we suspect it may be a new affective side effect of the thematic context for the sounds used in Nature.

\section{Discussion}

Although our results seem to suggest some trends with regards to our hypothesis, it should be noted that statistical analysis (pairwise one-way ANOVA) indicated that they were not sufficiently representative, so they should be taken with some caution.

The Nature ecology, for which annotation accuracy was the worst, seemed to be more positively received by testers in relation to the primary task, and to allow for a better performance at it. This may be related to a bias in tester demography, as the participants testing Nature were overall younger than those testing Synth (15 of them were aged below 40, against 8 for Synth), so they may just have been more proficient at mouse-based tasks in general. However, such an explanation does not account for the fact that the reaction times were overall shorter for Synth than for Nature. Perhaps testers for Synth and Nature simply prioritized different aspects of the dual task, respectively the annotation of stimuli or the copy of sequences. Still, the qualitative feedback seems to confirm the often-cited preference of listeners for natural-sounding stimuli, which may have transpired into their primary task performance as a sign of a more relaxed monitoring experience.

The fact that Synth was better detected but enjoyed less by testers may be due to the fact that it is closer to the traditional ``functional'' alarms paradigm, which allows for strong notifications but tend to be more stressful in long-term use.

We notice that some of the stimuli that were shared between ecologies were better detected when presented in the Mixed ecology than in their respective coherent ones: Arpeggio is more efficient in Mixed than Synth, and Sizzle and Water are more efficient in Mixed than Nature. This would suggest a confirmation of our original hypothesis for designing an ecology where different categories of sounds could better stand out from one another.

\section{Conclusion}
As part of our work towards the use of peripheral sonification in 3D printing process monitoring, we have studied the effect of ecological coherence in soundscapes. The comparative evaluation we conducted seems to favour our hypothesis that an incoherent ecology allows for better detection and discrimination of stimuli. Though, we also note that the natural-sounding ecology was seen as less intrusive by users, and that the synthetic-sounding one yielded shorter reaction times. When it comes to those criteria, our Mixed ecology's position between the other two seems to make it a sort of compromise for both low intrusion and fast reaction.

We notice that, although the Nature ecology performed overall worse than the others for stimulus discrimination, Birds did much better than Drone as a stimulus for PH notifications. Assuming incoherent ecology designs are worth researching further, we now wonder if it would be possible to build a better Mixed ecology by including that natural sound instead. More generally, in order to determine as exhaustively as possible which combination of stimuli allows for the best reaction rates, we would probably need to experiment on all of them. This would also allow us to test whether WPT and PT could be detected more accurately if they were conveyed by completely different sound types (Water + Bell, or Jingle + Sizzle).

For this work, we designed sets of stimuli that we deemed representative of coherent ecological considerations for either musical or natural sounds. Although the experiment we conducted seemed to favour the incoherent ecology, we carefully keep in mind that those results are still likely to reflect our own specific design choices rather than the types of sound ecologies themselves. So, we hope that a larger corpus of experiments comparing coherent and incoherent soundscape designs emerges in the future and allows for a more definite conclusion to be drawn on the topic of ecological incongruity in notification systems.

\begin{acknowledgments}
The authors of this paper would like to thank the SCRIME of the University of Bordeaux for hosting this research, as well as the Addimadour platform and Anaïs Do\-mergue for allowing us to observe the wire arc printing process in person, and answering our questions on the topic. We would also like to thank Nadine Couture and Sébastien Ibarboure (ESTIA) for initially submitting this case study to us and being greatly involved in the proof-reading and documentation of previous iterations of this work. Finally we want to acknowledge the contribution of Matthias Robine (LaBRI) and Emmanuel Duc (SIGMA Clermont) for their insights into sound design and choice of evaluation process.
\end{acknowledgments} 

\bibliography{smc2022bib}

\begin{thebibliography}{10}
\providecommand{\url}[1]{#1}
\csname url@samestyle\endcsname
\providecommand{\newblock}{\relax}
\providecommand{\bibinfo}[2]{#2}
\providecommand{\BIBentrySTDinterwordspacing}{\spaceskip=0pt\relax}
\providecommand{\BIBentryALTinterwordstretchfactor}{4}
\providecommand{\BIBentryALTinterwordspacing}{\spaceskip=\fontdimen2\font plus
\BIBentryALTinterwordstretchfactor\fontdimen3\font minus
  \fontdimen4\font\relax}
\providecommand{\BIBforeignlanguage}[2]{{%
\expandafter\ifx\csname l@#1\endcsname\relax
\typeout{** WARNING: IEEEtran.bst: No hyphenation pattern has been}%
\typeout{** loaded for the language `#1'. Using the pattern for}%
\typeout{** the default language instead.}%
\else
\language=\csname l@#1\endcsname
\fi
#2}}
\providecommand{\BIBdecl}{\relax}
\BIBdecl

\bibitem{kramer1999}
G.~Kramer, B.~N. Walker, T.~Bonebright, P.~Cook, J.~H. Flowers, and N.~Miner,
  ``The {Sonification} {Report}: {Status} of the {Field} and {Research}
  {Agenda}. {Report} prepared for the {National} {Science} {Foundation} by
  {Members} of the {International} {Community} for {Auditory} {Display},'' in
  \emph{International {Community} for {Auditory} {Display} ({ICAD})}, Santa Fe,
  NM, 1999.

\bibitem{gaver1991}
W.~W. Gaver, R.~B. Smith, and T.~O'Shea, ``{Effective Sounds in Complex
  Systems: The {ARKola} Simulation},'' in \emph{Proceedings of the SIGCHI
  Conference on Human factors in Computing Systems}, 1991, pp. 85--90.

\bibitem{rauterberg1994}
M.~Rauterberg and E.~Styger, ``{Positive Effects of Sound Feedback During the
  Operation of a Plant Simulator},'' in \emph{International Conference on
  Human-Computer Interaction}.\hskip 1em plus 0.5em minus 0.4em\relax Springer,
  1994, pp. 35--44.

\bibitem{patterson1990}
R.~D. Patterson, T.~F. Mayfield, D.~E. Broadbent, A.~D. Baddeley, and
  J.~Reason, ``Auditory {Warning} {Sounds} in the {Work} {Environment},''
  \emph{Philosophical Transactions of the Royal Society of London. B,
  Biological Sciences}, vol. 327, no. 1241, pp. 485--492, 1990.

\bibitem{weiser1996}
M.~Weiser and J.~S. Brown, ``The {Coming} {Age} of {Calm} {Technology},'' 1996.

\bibitem{vickers2011}
P.~Vickers, ``\BIBforeignlanguage{English}{Sonification for {Process}
  {Monitoring}},'' in \emph{\BIBforeignlanguage{English}{The Sonification
  Handbook}}, T.~Hermann, A.~Hunt, and J.~Nuehoff, Eds.\hskip 1em plus 0.5em
  minus 0.4em\relax Logos Verlag, 2011, ch.~18, pp. 455--491.

\bibitem{hildebrandt2016}
\BIBentryALTinterwordspacing
T.~Hildebrandt, T.~Hermann, and S.~Rinderle-Ma, ``{Continuous Sonification
  Enhances Adequacy of Interactions in Peripheral Process Monitoring},''
  \emph{International Journal of Human-Computer Studies}, vol.~95, pp. 54--65,
  2016. [Online]. Available:
  \url{https://www.sciencedirect.com/science/article/pii/S107158191630074X}
\BIBentrySTDinterwordspacing

\bibitem{mynatt1998}
E.~D. Mynatt, M.~Back, R.~Want, M.~Baer, and J.~B. Ellis, ``Designing {Audio}
  {Aura},'' in \emph{Proceedings of the SIGCHI conference on Human factors in
  computing systems}, 1998, pp. 566--573.

\bibitem{murray2006}
M.~M. Murray, C.~Camen, S.~L.~G. Andino, P.~Bovet, and S.~Clarke, ``Rapid
  {Brain} {Discrimination} of {Sounds} of {Objects},'' \emph{Journal of
  Neuroscience}, vol.~26, no.~4, pp. 1293--1302, 2006.

\bibitem{lomber2008}
S.~G. Lomber and S.~Malhotra, ``{Double Dissociation of 'What' and 'Where'
  Processing in Auditory Cortex},'' \emph{Nature neuroscience}, vol.~11, no.~5,
  pp. 609--616, 2008.

\bibitem{leaver2010}
A.~M. Leaver and J.~P. Rauschecker, ``Cortical {Representation} of {Natural}
  {Complex} {Sounds}: {Effects} of {Acoustic} {Features} and {Auditory}
  {Object} {Category},'' \emph{Journal of Neuroscience}, vol.~30, no.~22, pp.
  7604--7612, 2010.

\bibitem{isnard2016}
\BIBentryALTinterwordspacing
V.~Isnard, M.~Taffou, I.~Viaud-Delmon, and C.~Suied, ``Auditory {Sketches}:
  Very {Sparse} {Representations} of {Sounds} {Are} {Still} {Recognizable},''
  \emph{PLOS ONE}, vol.~11, no.~3, pp. 1--15, 03 2016. [Online]. Available:
  \url{https://doi.org/10.1371/journal.pone.0150313}
\BIBentrySTDinterwordspacing

\bibitem{poret2021}
M.~Poret, S.~Ibarboure, M.~Desainte-Catherine, C.~Semal, and N.~Couture,
  ``Peripheral {Auditory} {Display} for {3D-Printing} {Process} {Monitoring},''
  in \emph{Proceedings of the 2021 {Sound} {and} {Music} {Computing}
  {Conference} ({SMC} 2021)}, Turin, Italy, 2021.

\bibitem{cohen1993}
\BIBentryALTinterwordspacing
J.~Cohen, ``{"Kirk Here"}: Using {Genre} {Sounds} to {Monitor} {Background}
  {Activity},'' in \emph{INTERACT '93 and CHI '93 Conference Companion on Human
  Factors in Computing Systems}, ser. CHI '93.\hskip 1em plus 0.5em minus
  0.4em\relax New York, NY, USA: Association for Computing Machinery, 1993, p.
  63–64. [Online]. Available: \url{https://doi.org/10.1145/259964.260073}
\BIBentrySTDinterwordspacing

\bibitem{gilfix2000}
M.~Gilfix and A.~L. Couch, ``Peep ({The} {Network} {Auralizer}): {Monitoring}
  {Your} {Network} with {Sound},'' in \emph{Proceedings of the 14th {USENIX}
  {Conference} on {System} {Administration} ({LISA}'00)}.\hskip 1em plus 0.5em
  minus 0.4em\relax Berkeley, CA, USA: USENIX Association, 2000, pp. 109--118.

\bibitem{hermann2015}
T.~Hermann, T.~Hildebrandt, P.~Langeslag, and S.~Rinderle-Ma, ``Optimizing
  {Aesthetics} and {Precision} in {Sonification} for {Peripheral} {Process}
  {Monitoring},'' in \emph{Proceedings of the 21st {International} {Conference}
  for {Auditory} {Display} ({ICAD} 2015)}, Graz, Austria, 2015, pp. 317--318.

\bibitem{lenzi2019}
S.~Lenzi, T.~Riccardo, T.~Ginevra, S.~Galelli, P.~Ciuccarelli \emph{et~al.},
  ``{Disclosing Cyber-Attacks on Water Distribution Systems. An Experimental
  Approach to the Sonification of Threats and Anomalous Data},'' in \emph{25th
  International Conference on Auditory Display}.\hskip 1em plus 0.5em minus
  0.4em\relax International Community on Auditory Display, 2019, pp. 125--132.

\bibitem{aldana2020}
A.~L. Aldana~Blanco, S.~Grautoff, and T.~Hermann, ``{ECG Sonification to
  Support the Diagnosis and Monitoring of Myocardial Infarction},''
  \emph{Journal on Multimodal User Interfaces}, vol.~14, no.~2, pp. 207--218,
  2020.

\bibitem{matinfar2019}
S.~Matinfar, T.~Hermann, M.~Seibold, P.~F{\"u}rnstahl, M.~Farshad, and
  N.~Navab, ``Sonification for {Process} {Monitoring} in {Highly} {Sensitive}
  {Surgical} {Tasks},'' in \emph{Proceedings of the Nordic Sound and Music
  Computing Conference 2019 (Nordic SMC 2019)}, 2019.

\bibitem{williams2016}
\BIBentryALTinterwordspacing
S.~W. Williams, F.~Martina, A.~C. Addison, J.~Ding, G.~Pardal, and
  P.~Colegrove, ``Wire + {Arc} {Additive} {Manufacturing},'' \emph{Materials
  Science and Technology}, vol.~32, no.~7, pp. 641--647, 2016. [Online].
  Available: \url{https://doi.org/10.1179/1743284715Y.0000000073}
\BIBentrySTDinterwordspacing

\bibitem{ibarboure2021}
\BIBentryALTinterwordspacing
S.~Ibarboure, ``{Perception d'Information par des Retours Vibrotactiles pour
  une Mise en {\OE}uvre plus Flexible des Proc{\'e}d{\'e}s Robotis{\'e}s de
  Fabrication Additive.}'' Theses, {Universit{\'e} de Bordeaux}, Jun. 2021.
  [Online]. Available: \url{https://tel.archives-ouvertes.fr/tel-03361862}
\BIBentrySTDinterwordspacing

\bibitem{Ceruti2017}
A.~Ceruti, A.~Liverani, and T.~Bombardi, ``Augmented {Vision} and {Interactive}
  {Monitoring} in {3D} {Printing} {Process},'' in \emph{International Journal
  on Interactive Design and Manufacturing}, 2017.

\bibitem{poret2020}
M.~Poret, S.~Ibarboure, M.~Robine, E.~Duc, N.~Couture, M.~Desainte-Catherine,
  and C.~Semal, ``Sonification for {3D Printing} {Process} {Monitoring},'' in
  \emph{Proceedings of the 2021 International Computer Music Conference (ICMC
  2021)}, Santiago, Chile, 2021.

\bibitem{xia2020}
\BIBentryALTinterwordspacing
C.~Xia, Z.~Pan, J.~Polden, H.~Li, Y.~Xu, S.~Chen, and Y.~Zhang, ``A {Review} on
  {Wire} {Arc} {Additive} {Manufacturing}: Monitoring, {Control} and a
  {Framework} of {Automated} {System},'' \emph{Journal of Manufacturing
  Systems}, vol.~57, pp. 31--45, 2020. [Online]. Available:
  \url{https://www.sciencedirect.com/science/article/pii/S0278612520301412}
\BIBentrySTDinterwordspacing

\bibitem{gaver1993}
W.~W. Gaver, ``What in the {World} {Do} {We} {Hear}? an {Ecological} {Approach}
  to {Auditory} {Event} {Perception},'' \emph{Ecological Psychology}, vol.~5,
  no.~1, pp. 1--29, 1993.

\bibitem{blattner1989}
M.~M. Blattner, D.~A. Sumikawa, and R.~M. Greenberg, ``Earcons and {Icons}:
  Their {Structure} and {Common} {Design} {Principles},'' \emph{Human--Computer
  Interaction}, vol.~4, no.~1, pp. 11--44, 1989.

\bibitem{gaver1989}
W.~W. Gaver, ``The {SonicFinder}: An {Interface} that {Uses} {Auditory}
  {Icons},'' \emph{Human--Computer Interaction}, vol.~4, no.~1, pp. 67--94,
  1989.

\bibitem{macmillan2004}
N.~A. Macmillan and C.~D. Creelman, \emph{Detection {Theory}: A {User}'s
  {Guide}}.\hskip 1em plus 0.5em minus 0.4em\relax Psychology press, 2004.

\end{thebibliography}

\end{document}